\documentclass[10pt,conference,a4paper]{IEEEtran}

\usepackage{times,amsmath,epsfig,graphicx}
\usepackage{epstopdf}
\title{Convolutional Neural Network Steganalysis's Application to Steganography}
\author{%
{Mehdi Sharifzadeh{\small $~^{\#1}$}, Chirag Agarwal{\small $~^{\#2}$}, Mohammed Aloraini{\small $~^{\#3}$}, Dan Schonfeld{\small $~^{\#4}$}}%
\vspace{1.6mm}\\
\fontsize{10}{10}\selectfont\itshape
$^{\#}$\,ECE Department, University of Illinois at Chicago\\
Chicago, IL, USA\\
\fontsize{9}{9}\selectfont\ttfamily\upshape
\vspace{1.2mm}\\
$^{1}$\,mshari5, $^{2}$\,cagarw2, $^{3}$\,malora2, $^{4}$\,dans [@uic.edu]\\
\fontsize{10}{10}\selectfont\rmfamily\itshape
}
\usepackage{mathtools}
\usepackage{amsmath}
\usepackage{amssymb}
\usepackage{cite}
\DeclareMathOperator*{\argmin} {arg\,min}

\DeclareMathOperator*{\E} {E}
\DeclareMathOperator*{\entropy} {H}
\DeclareMathOperator*{\NN} {NN_o}
\DeclarePairedDelimiter\abs{\lvert}{\rvert}%
\begin{document}
\maketitle


\begin{abstract}
This paper presents a novel approach to increase the performance bounds of image steganography under the criteria of minimizing distortion. The proposed approach utilizes a steganalysis convolutional neural network (CNN) framework to understand an image's model and embed in less detectable regions to preserve the model. In other word, the trained steganalysis CNN is used to calculate derivatives of the statistical model of an image with respect to embedding changes. The experimental results show that the proposed algorithm outperforms previous state-of-the-art methods in a wide range of low relative payloads when compared with HUGO, S-UNIWARD, and HILL by the state-of-the-art steganalysis.
\\[1\baselineskip]
\end{abstract}
\section{Introduction}
\label{sec:intro}
Steganography is used for a wide range of application such as securely storing sensitive data, e.g. system passwords or keys, within other files, communication privacy by hiding its existence, and also watermarking application. Watermarking is important in its own way as it is used in copyright protection, source tracking, and authentication purposes, and etc.

In steganography problem,  modeled by the prisoner’s problem \cite{prisoner}, a hidden message is embedded by Alice in a cover medium with a private or public key, producing a stego medium from which the message should be decodable by Bob, without raising any suspicion from the warden, Wendy. Wendy can also alter the stego message to prevent any
communication 
 in case of being an active warden. However, we are only considering the case of passive warden, whose goal is to reveal the existence of any hidden message. Security of a steganography method is measured by how difficult the disclosing is for the warden or the steganalyzer, formally formulated in \cite{cachin1998information}, where Cachin defines perfectly secure and $\epsilon$-secure steganography. Further investigation from information theoretic point of view was done by Moulin et al. \cite{moulin2003information,moulin2007capacity}, where theoretic bounds were calculated for both cases which can be reached by cover generation methods \cite{ryabko2009asymptotically}. However, these approaches need accurate knowledge of the probability distribution of the cover media, which is not feasible for empirical non-stationary media such as images or videos.

In practical approaches, embedding is done while minimizing the caused distortion which can be formulated to a source coding problem with a fidelity criterion \cite{shannon1959coding}. One approach for minimizing the distortion is to make very small changes in spatial domain at the noise level. For example, one of the most popular image steganography methods is altering the least significant bit of pixels individually according to the hidden message bits \cite{cheddad2010digital,johnson1998exploring}. However, because of the dependent noise and pixel to pixel dependencies in images, it can be easily detected \cite{fridrich2001reliable}. So for achieving a better security, embedding should be done in more complex textures or noisy areas rather than smooth regions.

This has led to a group of spatial image steganography methods that we call cost based algorithms. They have two main steps, first is calculating the cost of embedding in each pixel using a suitably defined distortion function, second is embedding the message according to the costs. Second step is solved for a rather general class of distortion functions using syndrome trellis codes \cite{filler2011minimizing,filler2010gibbs}. As a result, the main focus in cost based image steganography is on the first step, deriving a cost function, for which we have proposed a new approach in Section \ref{ssec:From Steganlysis to Steganography}. In the second step, we will employ a method proposed in \cite{sharifzadeh2017arxiv} which outperforms the commonly used methodology for distributing the message among pixels.

Examples of cost based algorithms are HUGO \cite{HUGO}, S-UNIWARD \cite{S-UNIWARD} and HILL \cite{HILL}. These methods utilize different ways for distortion calculation. HUGO defines the distortion as a weighted sum of difference between SPAM feature vector of a cover image and its stego version \cite{pevny2010steganalysis}. In spatial universal wavelet relative distortion (S-UNIWARD) the embedding distortion is calculated using directional filter banks. HILL uses a high-pass filter to find noisy parts in an image, and then uses two low-pass filters to smooth the calculated costs. We utilize a steganalyzer, a trained CNN \cite{neuralnet}, to calculate pixel cost by measuring the network's output for different embedding patterns.  

This paper is organized as follows. We introduce notations and review the preliminaries and the problem formulation in Section \ref{sec:prior}. The proposed method is explained in Section \ref{method}. Results of comparative experiments are given in Section \ref{sec:experiments} to demonstrate the effectiveness of the proposed method. Conclusions are drawn in Section \ref{sec:conclusions}.
%
%
%
%
%
\section{Cost-Based Steganography}
\label{sec:prior}
\subsection{Notation}
\label{ssec:notation}
Capital and lower case bold face symbols are used for matrices and vectors respectively. $\entropy$ is the entropy function in bits, and all the logarithms are binary. $\E$ is the probability expectation function.
\subsection{Problem Formulation}
\label{ssec:formulation}
We use $\textbf{X}=(x_{ij})^{n_1 \times n_2} \in \mathcal{X}=\{0,\dots,255\}^{n_1 \times n_2}$, set of all 8-bit gray-scale images of size $n_1 \times n_2$,  for cover image. By assuming a ternary embedding scenario, $\textbf{Y}=(y_{ij})^{n_1 \times n_2}$ is the corresponding stego image, where $y_{ij} \in \mathcal{I}_{ij}=\{\max(0,x_{ij}-1),x_{ij},\min(255,x_{ij}+1)\}$. As a result, the embedding pattern can be defined as $\textbf{S}=\textbf{Y}-\textbf{X}=(s_{ij})^{n_1 \times n_2} \in \mathcal{S}=\{-1,0,+1\}^{n_1 \times n_2}$, which is the coded stego message using practical coding schemes as syndrome trellis codes \cite{filler2011minimizing}, and it is chosen according to the probability distribution $p(\textbf{S})$. 

In case of having a fixed relative payload, total number of bits over the total number of pixels, the goal is first to define a distance function $D( \textbf{X}, \textbf{Y}): \mathcal{X} \times \mathcal{X} \to \mathbb{R}$ for calculating embedding impact, then to solve the constrained optimization problem below using the suitably defined function $D$. 

\begin{equation} \label{main}
\argmin_{p} \E(D(\textbf{X},\textbf{Y})) = \argmin_{p} \sum_{\textbf{S} \in \mathcal{S}}{D(\textbf{X},\textbf{Y})\times p(\textbf{S})}
\end{equation}
\begin{equation} \label{message constraint}
m = \entropy(p) \triangleq -\sum_{\textbf{S} \in \mathcal{S}}{p(\textbf{S})\times \log\big(p(\textbf{S})\big)}
\end{equation}
where $m$ is the length of the message in bits.

Below is the solution of this problem using Lagrangian multipliers method \cite{fridrich2007practical}:
\begin{equation} \label{exponential}
p(\textbf{S}) = \frac{e^{-\lambda D(\textbf{S})}}{\sum_{\textbf{S} \in \mathcal{S}}{e^{-\lambda D(\textbf{S})}}}
\end{equation}
where $\lambda$ is the Lagrangian multiplier which can be determined from (\ref{message constraint}). By assuming mutually independent and symmetrical embedding impacts under an additive distortion scenario, a reasonable assumption also made in HUGO, HILL and S-UNIWARD, distance function can be written as below:
\begin{equation} \label{additive distortion}
D(\textbf{X},\textbf{Y}) = \sum\limits_{i=1}^{n_1} \sum\limits_{j=1}^{n_2}
 \rho_{ij} \abs{x_{ij}-y_{ij}}
\end{equation}
where $\rho_{ij}$ is the embedding impact of changing only one pixel by $\pm1$ which is called the cost of embedding. This will result in the following probability distribution for embedding changes:
\begin{equation} \label{additive probability}
p(s_{ij}) = 
  \begin{cases} 
   \frac{e^{-\lambda \rho_{ij} }}{1+2e^{-\lambda  \rho_{ij}}} & \text{if  $s_{ij} = \pm 1$} \\
   \frac{1}{1+2e^{-\lambda  \rho_{ij}}}       & \text{if  $s_{ij} = 0$}
  \end{cases}
\end{equation}
The main design principle of the proposed method is using steganalysis for steganography and calculating pixel costs($\rho_{ij}$). 
\subsection{Steganalysis-Based Steganography}
\label{ssec:From Steganlysis to Steganography}
Almost all state-of-the-art statistical steganalyzers (except steganalyzers for LSB replacement) are based on calculating steganalytic features and feeding them to a machine learning algorithms. In steganalysis, steganalytic features are designed to be more sensitive to embedding changes rather than differences among covers. As a result, the learning algorithm will be able to learn how the features will change in case of embedding message regardless of the changes in features for different covers. 
So from steganography point of view, the features can serve as an image model to determine the embedding costs $\rho_{ij}$. This transition from steganalysis to a steganographic model can be used with different steganalysis methods and in different domains as well.

\begin{figure}[t]
\begin{minipage}[b]{1.\linewidth}
  \centerline{\includegraphics[width=8cm]{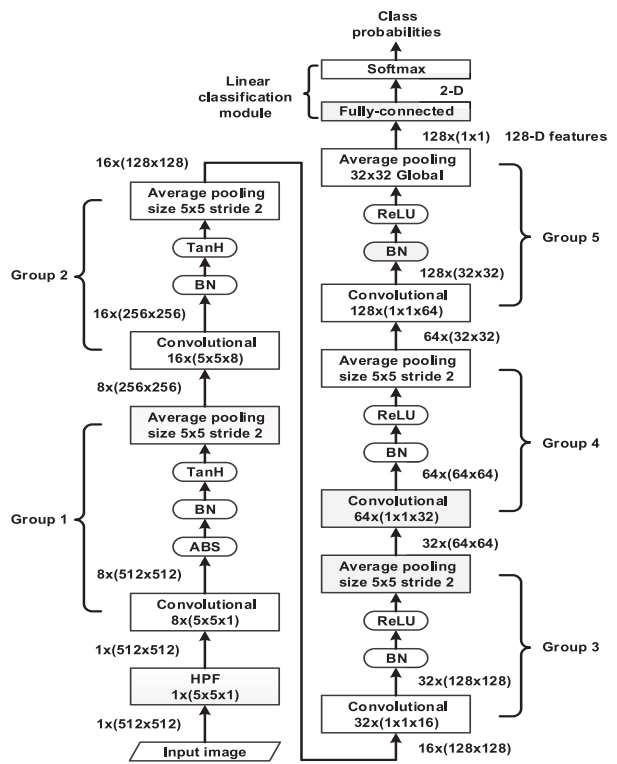}}
\end{minipage}
\caption{Convolutional neural network architecture \cite{neuralnet}}
\label{architecture}
\end{figure}

This idea has been previously done using SPAM features \cite{SPAM} having a dimensionality of 686 in HUGO \cite{HUGO}. We believe that using higher dimensional features will result in better security than HUGO but it will be time consuming given the fact that even HUGO is a very slow algorithm. One of the best known steganalytic features are Spatial Rich Model features (SRM) having a dimensionality of 34,671\cite{SRM}. Using these features directly for developing an steganography algorithm is almost impossible, given the size of these features, as they have to be calculated for numerous different embedding patterns for each cover media. Thus instead of using these features, we are proposing using another steganlysis algorithm which makes this task computationally feasible. The steganalysis method uses a deep CNN shown in Figure \ref{architecture} proposed by Xu et al. \cite{neuralnet} and it outperforms the steganalysis using ensemble classifier trained on SRM features \cite{classifier}. After training the CNN, the probabilty of being an stego message is produced very fast for each input image. As a result, this probability can be calculated many times for different embedding patterns in each cover in a feasible time to estimate the embedding costs $\rho_{ij}$. This will be explained in details in the following section.

\section{Methodology}
\label{method}
Lets assume that $\textbf{X}_{ij^{+d}}$, and $\textbf{X}_{ij^{-d}}$ are the altered cover images in which $x_{ij}$ is changed by $+d$, and $-d$ respectively. By having the assumption of mutually independent embedding impacts under an additive distortion scenario, $\rho$ can be derived for each pixel separately as the sensitivity of the image model learned by CNN to changes in each pixel. It is to be noted that the assumption gets highly violated once we start increasing the payload. The sensitivity is estimated by second derivative of the CNN's output probability with respect to the pixel value change.
\begin{equation} \label{network derivative}
\rho_{ij} = \max(\frac{\partial^2 \NN(\textbf{X}) }{\partial x_{ij}^2},0)
\end{equation}
Where $\NN$, the output of the trained CNN, is the probability of the image being a stego message. For estimating the second derivative, the following finite difference formula with $4^{th}$ order accuracy is used\cite{derivative}:
\begin{equation}\label{derivative est}
\begin{aligned}
\frac{\partial^2 \NN(\textbf{X}) }{\partial x_{ij}^2}&\approx\frac{-\NN(\textbf{X}_{ij^{-2}})}{12}+\frac{4\NN(\textbf{X}_{ij^{-1}})}{3}\\
&~+\frac{-5\NN(\textbf{X}_{ij})}{2}+\frac{4\NN(\textbf{X}_{ij^{+1}})}{3}\\
&~+\frac{-\NN(\textbf{X}_{ij^{+2}})}{12}
\end{aligned}
\end{equation}
The result of this step is then transformed to $\rho$'s by scaling using linear mapping, and smoothing using an average filter. Average filter also helps in compensating for the loss due to the assumption of independent embedding impacts. Note that the authors of HILL also used an average filter for smoothing and it improved their algorithm using a certain size. Different average filters with different sizes are tested and the results are provided in Table \ref{filters} which will be discussed in the next section. In addition to scaling and smoothing, embedding in saturated pixels, pixels with 0 and 255 intensities, is totally avoided by setting the corresponding $\rho$'s to a very high cost \cite{saturatedpixels}. By using the calculated $\rho$'s, although (\ref{additive probability}) is proven to be the solution of (\ref{main}), embedding probabilities are driven using the following formula instead of (\ref{additive probability}), as it is shown to have a better performance in \cite{sharifzadeh2017arxiv}.
\begin{equation} \label{new}
p(s_{ij}) = 
  \begin{cases} 
   \max(\frac{1}{3}-\lambda \rho_{ij},0) & \text{if  $s_{ij} = \pm 1$} \\
   1-2\max(\frac{1}{3}-\lambda \rho_{ij},0)       & \text{if  $s_{ij} = 0$}
  \end{cases}
\end{equation}
Using this framework for calculating pixel costs and embedding probabilities allows us to “preserve” the model utilized by steganalysis and thus embed in a more secure way comparing to other spatial image steganography algorithms.
\section{Experiments}
\label{sec:experiments}
\begin{table}[t]
\centering
\caption{\textrm{\normalfont Out-of-Bag error (detection error) for payload=0.4 (bits per pixel) of the proposed algorithm using different average filter sizes.}}
\renewcommand{\arraystretch}{1.5}
\label{filters}
\begin{tabular}{|c|l|l|l|l|l|}
\hline
Average Filter & $1\times1$ & $3\times3$ & $7\times7$ & $13\times13$ & $21\times21$ \\ \hline
$E_{OOB}$&0.1821&0.2374& 0.3091& \textbf{0.3439}&0.3265\\ \hline
\end{tabular}
\end{table}
\subsection{Main Configuration}
\label{main setup}
Throughout this paper, all the experiments are conducted on BOSSbase ver.1.01 database \cite{BOSSbase} containing 10,000 gray-scale $512\times512$ pixels images. 
Performance evaluations are done by an ensemble classifier steganalyzer \cite{classifier} with a 10-fold cross validation trained on 34,671 dimensional SRM feature set. 5000 images are selected randomly in each experiment for training/validation and 1000 images are selected from the rest of the database for testing. Out-of-Bag Error is reported which is the average false positive and negative rates in testing step.

\subsection{Prior Work's Configuration}
For comparing the security of the proposed method, we have used three state-of-the-art spatial steganography algorithms with their best setting. HUGO is used with the same setting reported in the original paper. S-UNIWARD algorithm is used with $\sigma=1$, shown to be optimum in \cite{S-UNIWARDsigma}. HILL algorithm is used with a $3\times3$ KerBohme high-pass filter and $15\times15$ averaging filters as low pass filters.

\subsection{Convolutional Neural Network Configuration}
The CNN used for calculating pixel costs is trained on 10000 images of BOSSbase ver.1.01 database \cite{BOSSbase} and their stego versions using S-UNIWARD and HILL algortihm with relative payload of 0.1, 0.4 (2500 images each, chosen randomly without overlap). 8000 image pairs are used for training and 2000 for validation. Every image and its stego version are kept in the same batch, thus forming 64 images per batch. The network has the same configuration and hyper parameters as proposed in the original paper \cite{neuralnet}. After training step, the network has reached the validation accuracy of 69.81 percent. Then the trained network is used to calculate cost matrices for 6000 images, and embedding is done using these matrices for different relative payloads. Performance evaluation is done for each payload as it is explained in \ref{main setup}.

\subsection{Average Filter Size}
Different average filter sizes are used for smoothing the derivatives and transforming them to pixel costs. As it is explained in Section \ref{method}, averaging is done to smooth the costs and compensate the violation of assuming mutually independent costs in practice. The result of using different average filters are compared in Table \ref{filters}, which shows $13\times13$ average filter results in the highest security.

\begin{figure}[t]
\begin{minipage}[b]{1.0\linewidth}
  \centerline{\includegraphics[width=9.5cm]{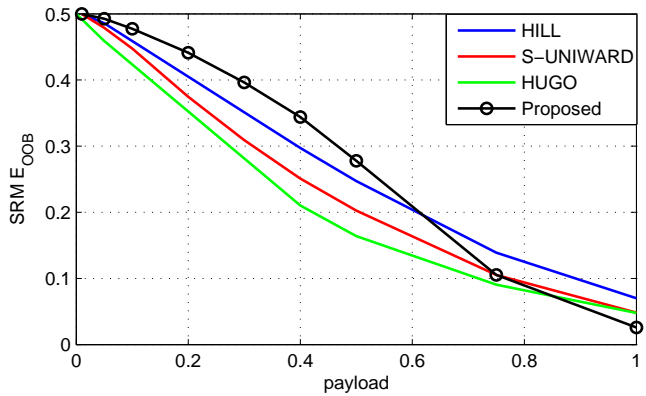}}
\end{minipage}
\caption{Out-of-Bag error (detection error) versus relative payload (bits per pixel) of HILL, S-UNIWARD, HUGO and the proposed algorithms applied on BOSSbase database ver. 1.01 calculated by an ensemble classifier trained on SRM features.}
\label{all vs neural}
\end{figure}
\subsection{Analysis}
\label{result analysis}
The comparison results are shown in Figure \ref{all vs neural} which shows higher performance of the proposed method comparing to three other algorithms up to relative payload of 0.6 bits per pixels. For higher payloads we observe performance drop. We believe it is the result of violation of the assumption of having mutually independent embedding impacts which we assumed in both pixel cost calculation and embedding step. This violation is more when the payload is higher because of having a dense embedding pattern matrix $\textbf{S}$ rather than a sparse matrix which we have in low payloads. However the other three methods only have this assumption in the second step for deriving (\ref{additive probability}).

The results are also consistent with this statement as it can be seen in Figure \ref{all vs neural} that the proposed algorithm performs better for a wide range of low payloads (0-0.62 bits per pixel) but worse in high payloads (greater than 0.62 bits per pixel) comparing to other algorithms.

\section{Conclusions}
\label{sec:conclusions}
A spatial image steganography method is presented which outperforms the state-of-the-art algorithms in certain payloads (0-0.62 bits per pixel). In this methodology, pixel costs are calculated utilizing a trained steganalysis convolutional neural network. Cost matrix is estimated by measuring the sensitivity of the output of the network to each pixel's intensity by assuming mutually independent embedding impacts. Although we compensate for the violation of this assumption in practice, it causes low performance in high payloads (greater than 0.62 bits per pixel). We will try to find better compensation strategies than average filtering in future. Another interesting future work would be utilizing a more complex embedding pattern than changing the image pixel by pixel for measuring the embedding distortions which can also help in compensation.
\pagebreak
\bibliographystyle{IEEEtran}
\bibliography{refs}

\enlargethispage{-5mm}
\end{document}